\documentstyle[12pt,epsf,psfig]{article}
 
\def\be{\begin{equation}}
\def\bea{\begin{eqnarray}}
\def\ee{\end{equation}}
\def\eea{\end{eqnarray}}

\begin{document}
 
\title{Conductivity Exponent and Backbone Dimension in 2-d Percolation}
\author{Peter Grassberger\\
{\sl \small HLRZ, Forschungszentrum J\"ulich, D-52425 J\"ulich, Germany}}
\maketitle
 
\medskip
\medskip
 
\begin{abstract}
\noindent
We present high statistics simulations for 2-d percolation 
clusters in the ``bus bar" geometry at the critical point, 
for site and for bond percolation.
We measured their backbone sizes and electrical conductivities. 
For all sets of measurements we 
find large corrections to scaling, most of which do not seem to be described by 
single powers. Using single power terms for the corrections to 
scaling of the backbone masses, we would obtain fractal dimensions 
which are different for site and bond percolation, while the correct 
result is $D_b = 1.6432\pm 0.0008$ for both.
For the conductivity, the corrections to scaling are 
strongly non-monotonic for bond percolation. The exponent $t' = t/\nu$ 
is measured as $0.9826 \pm 0.0008$, in disagreement with the Alexander-Orbach 
and other conjectures.
\end{abstract}

\section{Introduction}

The critical behavior of percolation is quite well understood, mainly 
because of its relationship to the one-state Potts model \cite{stauffer}.
Due to this relationship and conformal invariance, all ``thermal" exponents 
are exactly known in $d=2$. There are however a number of critical exponents 
for percolation which have no thermal analogue. The most important of these are 
the backbone dimension $D_b$ and the conductivity exponent $t$ which is defined 
by $\sigma \sim (p-p_c)^t$ above the critical point. Here and in the following we 
assume a `bus bar' geometry of size $L\times L$, and $\sigma$ is the global conductivity, 
with each empty site/bond being an insulator and each occupied site/bond 
having a constant finite resistivity. 
In a bus bar geometry two opposite sides of a quadratic lattice are connected 
to superconductors, while the other two sides are open. The backbone is defined 
as the set of lattice sites/bonds which are connected to both bus bars through
mutually non-intersecting paths. Related to $D_b$ and $t$ are 
various exponents for conduction exactly at the critical point, for conductivity of 
conductor/superconductor mixtures, and for random walks 
starting either on the infinite incipient 
cluster, on the backbone, or at arbitrary lattice sites \cite{havlin}. Instead 
of $t$ we shall in the following discuss mostly $t'=t\nu$, with $\nu=4/3$ being the 
correlation length exponent. Exactly at threshold, the conductivity of a 
lattice of size $L$ scales as $\sigma \sim L^{-t'}$. At 
present, the most precise Monte Carlo simulations give $t' = 0.9745\pm 0.0015$ 
\cite{normand} and $D_b = 1.647 \pm 0.004$ \cite{grass}. These seem to be 
more precise than estimates based on exact enumerations \cite{essam} 
or other numerical methods.

Theoretical attempts to compute these ``athermal" exponents have turned out 
to be less easy. There have been numerous conjectures, most of them rather 
ad hoc. The best known is due to Alexander and Orbach \cite{alexander}, and 
gives in 2 dimensions $t' = 91/96 = 0.94791\ldots$. It was for some time 
considered as very accurate, but if we accept the above 
numerical estimate, it is ruled out by about 18 standard deviations.

It seems that there are no conjectures for $t'$ based on conformal invariance, 
but there are several conjectures for $D_b$. 
In two independent attempts, Larsson \cite{larsson} 
and Saleur \cite{saleur} obtained the value $D_b=25/16 = 1.5625$ which is 
obviously in serious disagreement with the above numerical estimate (and with 
other recent numerical estimates \cite{rintoul1,moukarzel}). 
More recently, Huber \cite{huber} obtained $D_b=79/48 = 1.64587$. This 
is in excellent agreement with the simulations, but its theoretical 
justification is not very clear.

The present investigation was started for several reasons. The first is that 
the random number generator used in \cite{grass} (the Stoll-Kirkpatrick 
generator R250 \cite{stoll}) had obviously created some problems. 
For $L\to\infty$ the spanning probability should tend to 1/2 in
the square bus bar geometry, with corrections of size 
$1/L$ \cite{ziff,aharoni}. While the convergence to 1/2
was seen in \cite{grass}, it was much slower than it should have been. As a
consequence, the value of $D_b$ quoted above should be taken with some caution.
Similar problems with R250 were found in other depth-first algorithms
\cite{landau,grass2}. The spreading simulations presented in sec.2 of \cite{grass}
should be unaffected by these problems since they used a breadth-first algorithm 
for which R250 seems to be safe.

Secondly, we considered the situation concerning the conductivity exponent as 
unsatisfactory. The random number generators used in \cite{normand} were 
obviously even worse than R250. The data for site percolation presented in table 1 of  
\cite{normand} show fluctuations which are much larger than the supposed 
statistical errors, and could hardly be blamed on corrections to scaling, 
because of their irregular behavior. In addition, the estimates obtained in 
\cite{normand} for site and bond percolation are not compatible within their 
error bars.

Finally, it seemed that many of the papers in the recent literature used quite 
slow algorithms, while fast algorithms have been applied only with moderate 
statistics. This is particularly true for conductivity. There, the fastest 
algorithm by far (for the square lattice!) seems to be the one by Lobb and 
Frank \cite{lobb,frank}. Its time complexity is roughly $L^2\log L$. This 
should be compared to the time complexity $L^3\times N$ for strips of size 
$L\times N$ (with $N\gg L$) for the algorithm used in \cite{normand}. 
Nevertheless, while several months of CPU time on a special purpose computer 
were used in \cite{normand}, only a few hours on an IBM 3081 mainframe 
were spent on the published data for the Lobb-Frank algorithm.
\footnote{Dr. Frank has, however, produced a substantial amount of yet unpublished 
data \cite{frank-unpub}. I am greatly indebted to him for sending me these data. 
They were extremely helpful in convincing me that my own simulations were 
correct.} In addition, there exists a vast literature in which 
random walks were used to estimate $t$ via the Einstein relation \cite{havlin}. I 
implemented several variants (walks on percolating clusters only, walks on 
backbones only, walks on all sites). Even 
though I spent substantial amounts of CPU time on them, results were 
extremely poor compared to the results described below.

Similarly, although the algorithm used for backbone identification in 
\cite{grass} has a complexity $L^{D_f}$ where $D_f=91/48=1.895...$, much slower 
algorithms are still in use. This seems particularly true for various versions 
of the burning algorithm \cite{stanley1,stanley2,roux,bunde,rintoul1,rintoul2}.
An algorithm with complexity slightly larger than $L^2$ was used recently 
in \cite{moukarzel,moukarzel2}. It should be pointed out that the algorithm 
of \cite{grass} works only for strictly planar graphs (no such restriction 
holds for burning and for the algorithms of \cite{moukarzel,moukarzel2}), 
but a slightly more complex algorithm with essentially
the same asymptotic behavior was given long ago by Tarjan \cite{tarjan} for 
arbitrary graphs (Tarjan's algorithm finds the `biconnected' part of any graph 
which for percolation is just the backbone). 

Indeed, the fastest algorithm for estimating conductivity turned out to be 
a combination of the Lobb-Frank algorithm with the backbone-identification 
algorithm of \cite{grass}. First the backbone is found (time $L^{D_f}$), and 
then the Lobb-Frank algorithm is run on the backbone. The time for the latter 
is still $\geq O(L^2)$, not $L^{D_b}\log L$, as one might have 
expected naively, and one has to add the CPU time needed for finding the 
backbone. Nevertheless, total absolute CPU times are reduced by roughly a factor 4
for the largest values of $L$ studied here.

In retrospect, however, the main result of the paper is something 
unexpected, namely the very strange behavior of the corrections to scaling.
Field theory predicts that the leading corrections are power behaved, 
with universal exponents derivable by means of renormalization group techniques. 
As a consequence it is often assumed that a single power behaved correction 
term is sufficient to describe all deviations from the asymptotic scaling 
laws. In many cases this is correct, in particular if the data which are 
to be interpreted are not too precise. But there are counter examples. 
The best known maybe is the spherical model where Luck \cite{luck} showed 
analytically that corrections to scaling are non-monotonic. Another example is 
spreading of 3-d percolation where Monte Carlo simulations would give 
non-universal critical exponents if these simulations were fitted by 
simple power-behaved corrections \cite{grass-percol3}.

The corrections to scaling discussed below seem even more strange, in particular 
those for the conductivity. A priori one should expect that bond percolation is 
better behaved than site percolation \cite{lobb,frank}. The first reason 
is that the critical point is exactly known to be $p_c=1/2$ \cite{stauffer}. 
Secondly, if the geometry is properly chosen (together with the bus bars, 
the lattice has shape $L\times (L+1)$ \cite{lobb}), duality shows that 
the spanning probability (i.e. the probability that the backbone is 
non-empty and that $\sigma>0$) is exactly 1/2 at $p=p_c$, for all $L$. Indeed, 
corrections to scaling are smaller for bond percolation than for site 
percolation. But while they are reasonably well described by a single 
power for the latter, they show at least one full oscillation for bond percolation. 
Indeed, from the bond percolation data alone we cannot exclude the possibility 
that they continue to oscillate with $\log L$ (such log-periodic oscillations do occur 
in a similar conduction problem for large applied voltage \cite{stauffer-sornette}). 
It is only the comparison with site percolation for very large $L$ which 
makes ongoing oscillations unlikely.  

In the next section we will discuss backbones, in sec.3 conductivities. 
Our conclusions will be drawn in sec.4.

\section{Backbones}

The backbone of a (site) percolation cluster which 
spans between two sets $\cal{A}$ and $\cal{B}$ of sites is defined as the 
set of those sites which are connected to both $\cal{A}$ and $\cal{B}$ by 
two mutually non-intersecting paths. Except for `Wheatstone bridge' type 
configurations, these are also those sites which carry non-zero current 
if the lattice is made of a finite resistance conductor, and a potential 
difference is applied between $\cal{A}$ and $\cal{B}$. In the geometry 
used in the present paper, $\cal{A}$ and $\cal{B}$ are the bus bars which 
are supposed to have infinite conductivity.

The backbone is essentially (if an additional bond is added between 
$\cal{A}$ and $\cal{B}$) what is called the set of {\it biconnected nodes} in 
computer science. For general graphs, the fastest known algorithm to generate it 
is a recursive depth-first algorithm \cite{tarjan} which runs in time 
$O(N)$ for a graph of $N$ nodes, provided the number of edges meeting at each 
node is finite. Nearly the same $N$-dependence was reached 
recently in a novel algorithm \cite{moukarzel} which can also be extended to 
rigidity backbones \cite{moukarzel2}. 
At least for large $N$ this is much faster 
than the `burning' type algorithms used traditionally by physicists 
\cite{stanley1,stanley2,roux,bunde,rintoul1,rintoul2}.
For strictly planar graphs, an even faster algorithm was given in \cite{grass} and 
applied to 2-d percolation in the `bus bar' geometry (this algorithm cannot be 
applied, e.g., to 2-d percolation with periodic lateral boundary conditions, since 
this is not a strictly planar graph). This algorithm has the same asymptotic 
complexity as Tarjan's algorithm, but it is roughly twice as fast and uses 
about half of the memory, since it needs one data structure less and needs 
only one pass through all sites, instead of two passes in Tarjan's algorithm.

For site percolation we used essentially the same algorithm as in \cite{grass}, 
where also a detailed description can be found. The main difference is that we 
realized that we do not have to erase disorder configurations (i.e., set all sites variables 
again to some default value) before starting a new configuration. Instead, we attach 
a long integer $s_i$ to each site, and initialize all $s_i$ to zero before starting 
the simulations. While building the first configuration, we consider a 
site as untested if $s_i <s_i^{(1)}$, and increase $s_i$ after testing it in the 
same way as in \cite{grass}, except that the information whether the actual 
path is upward or downward is no longer encoded in the sign of $s_i$ (as in \cite{grass}),
but in an additional array. After completing the $k$-th configuration, $s_i^{(k+1)}$ 
was set equal to $\max_i s_i$, and the process was repeated.
For bond percolation the algorithm is slightly more complicated, but is a 
rather straight forward modification.
For both cases, the algorithms were carefully checked by 
visually inspecting a large number of examples. Another test consisted in 
comparing the conductivity of the backbone with that of the entire lattice. The 
two should be identical, and were so in all tested cases.

For site percolation we place the bus bars at the $x$-axis and at the line $y=L+1$, 
and use as `wettable' lattice sites those with $0< x,y\le L$. If the backbones 
defined in this way contained $N$ sites, their dimension is defined as 
\be 
   \langle N\rangle \sim L^{D_b}\;.          \label{bb}
\ee
For bond percolation we place the bus bars at $y=0$ and at $y=L$. Horizontal 
wettable bonds are those with $1< x <L,\; 0<y<L$, and vertical wettable bonds 
are those with $1\le x \le L,\; 0<y<L$. The backbone can now be defined again 
as the set of biconnected sites, or as the set of biconnected bonds. We call 
the cardinalities of these sets $N_s$ and $N_b$. For both of them we expect 
the scaling law eq.(\ref{bb}), with the same value of $D_b$.

As random number generators we used the multiplicative rule $i_{n+1} = i_n*13^{13}+1\;
({\rm mod}\; 2^{63})$ and the four-tap shift-register generator of Ziff \cite{four-tap} 
with period $2^{9689}-1$. 
Both gave results in perfect agreement. For bond percolation at $p=1/2$ the
spanning probability was 1/2 for all values of $L$. For site percolation we 
used the value $p_c=0.592746$ found in \cite{ziff}. The spanning probability 
now converged to 1/2 with a power close to $-1$, in perfect agreement with \cite{ziff}. 
In order to check for systematic errors due to a wrong estimate of $p_c$, we made 
also runs at $p=0.59276$ which is about 30 standard errors larger than the central 
value of \cite{ziff}. These runs showed that the estimates given below should still be
correct even if the errors in $p_c$ were underestimated in \cite{ziff} by a factor of 2.

For site percolation, raw data including sample sizes and CPU times are given 
in table 1.  
Lattice sizes studied were powers of 2, with $L$ up to 4096. In addition we simulated 
lattices with $L=3,6,12,\ldots 96$. For each $L\leq 48$ the number of configurations 
was $>10^8$. This number decreased for larger $L$, see table 1.
All simulations were done on Sun Ultra and DEC Alpha work stations, but for convenience 
CPU times are quoted only for the DEC Aplha machines. Determining the backbone 
(and the entire spanning cluster) on a $4096\times 4096$  took in average less than 3 seconds. 
For larger $L$ the CPU time increased slightly faster (roughly $\sim L^2$), 
probably due to increasingly frequent cache misses.

For bond percolation we used similar lattice sizes and went up to similar statistics. 
Details are given in table 2. CPU times were not measured separately, as all these runs were
made together with conductivity measurements, but they should be similar to those 
for site percolation.

In order to obtain estimates for $D_b$, we computed effective dimensions according 
to 
\be
   D_{b,{\rm eff}}(L) = {\log[N(2L)/N(L/2)] \over \log 4}\;.       \label{deff}
\ee
Here $N$ stands for the number of sites in site percolation, for the number of sites 
in bond percolation, or for the number of bonds in bond percolation. This gives 
us three sequences of effective dimensions. 

\begin{figure}[ht]
   \centerline{\psfig{file=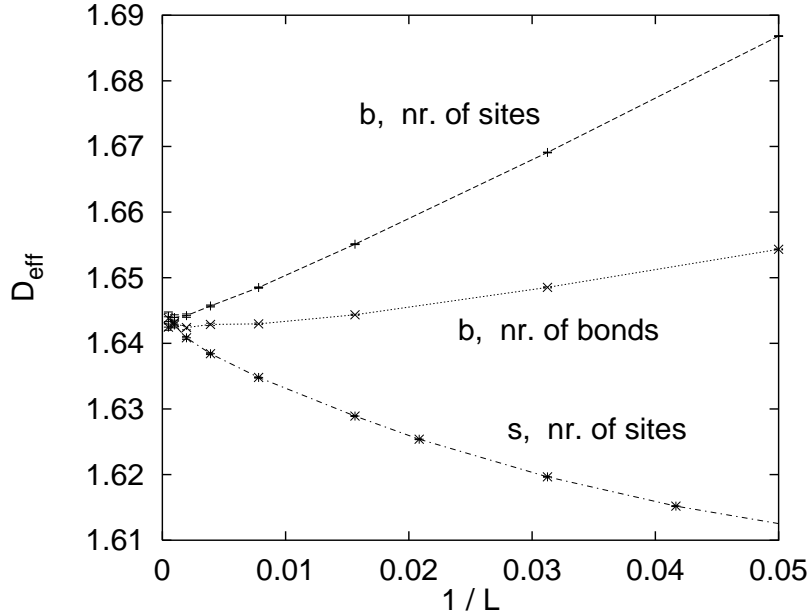,width=11cm,angle=270}}
   \caption{ \small 
        Effective backbone dimensions, estimated according to eq.(\ref{deff}), 
        plotted against $1/L$. The lowest curve is for site percolation, the uppermost
        for sites belonging to bond percolation backbones, and the middle for 
        bonds in bond percolation backbones. When error bars are not visible, they 
        are smaller than the sizes of the symbols.
   }
   \label{fig1}
\end{figure}

To obtain a first rough impression, we plot them in fig.1 against $1/L$. Only 
values for $L\geq 20$ are shown. All three curves 
seem to converge to the same value $\approx 1.643$, but with very different 
corrections to scaling. 

\begin{figure}[ht]
   \centerline{\psfig{file=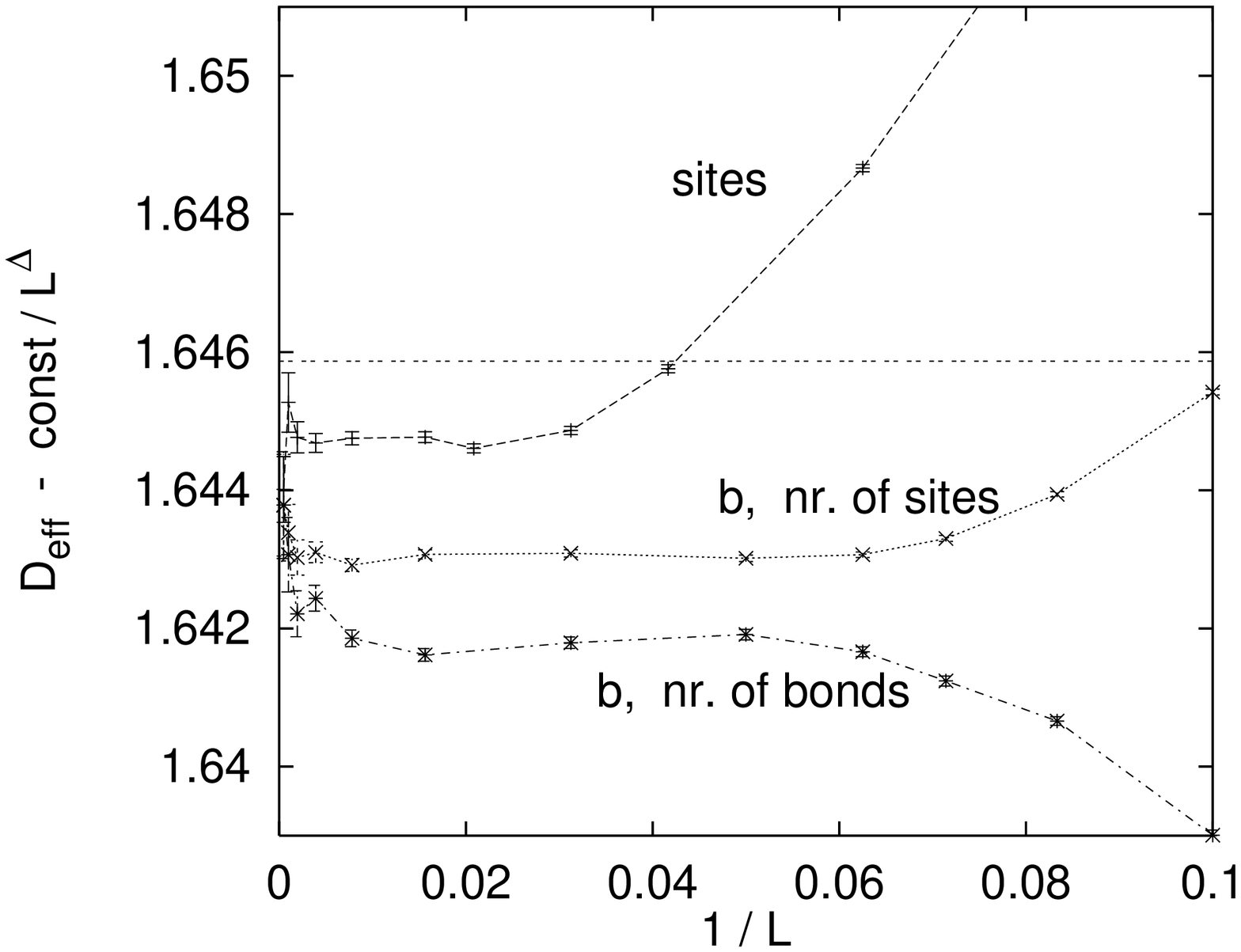,width=11cm,angle=270}}
   \caption{ \small 
        Effective backbone dimensions, after subtracting terms $a/L^\Delta$, 
        with $a$ and $\Delta$ obtained by fitting $D_{b,{\rm eff}}$ to ansatzes 
        $D_{b,{\rm eff}}= D_b + a/L^\Delta$ (these fits are done separately for 
        each of the three curves of fig.1). Notice that the range on the $y$-axis 
        has been reduced substantially, but the three curves do not 
        collapse. In particular, they do not extrapolate to the same value for 
        $L\to\infty$. The dashed horizontal line is the prediction of \cite{huber}.
   }
   \label{fig2}
\end{figure}

For site percolation (lowest curve) we see the strongest corrections. Here 
$D_{b,{\rm eff}}$ increases strongly with $L$, explaining why
$D_b$ was underestimated in the first analyses \cite{stanley1,stanley2}. 
The data seem to follow a smooth line, suggesting a single dominant correction 
term. Plotting the data against $1/L^\Delta$ and looking for a straight line does not 
produce entirely satisfactory results (showing that more than one term is needed), 
but if one insists in a single term, its power is $\Delta \approx 0.67$. The 
quality of this fit can be judged from fig.2 which shows the effective 
exponent after subtracting a term $a/L^\Delta$ with $\Delta=0.67$ and $a$ 
such that the data are fitted for $L\geq 32$. Notice that we show a larger 
range of $L$ in fig.2, and the fit with a single correction term becomes indeed 
bad for $L\le 24$.

The same kind of analysis was also done for bond percolation. Again we see 
that the data plotted against $1/L^\Delta$ do not give perfect lines for any $\Delta$, 
but the best fitting values are definitely larger than for site percolation: 
$\Delta = 1.11$ for $N_s$, and $\Delta = 1.30$ for $N_b$. As expected, it is 
bond percolation with $N_b$ defining the backbone mass which gives the largest 
$\Delta$ and, as seen from fig.1, also the smallest amplitude for the correction 
to scaling. The results obtained by subtracting the fitted correction terms 
are also plotted in fig.2.

\begin{figure}[ht]
   \centerline{\psfig{file=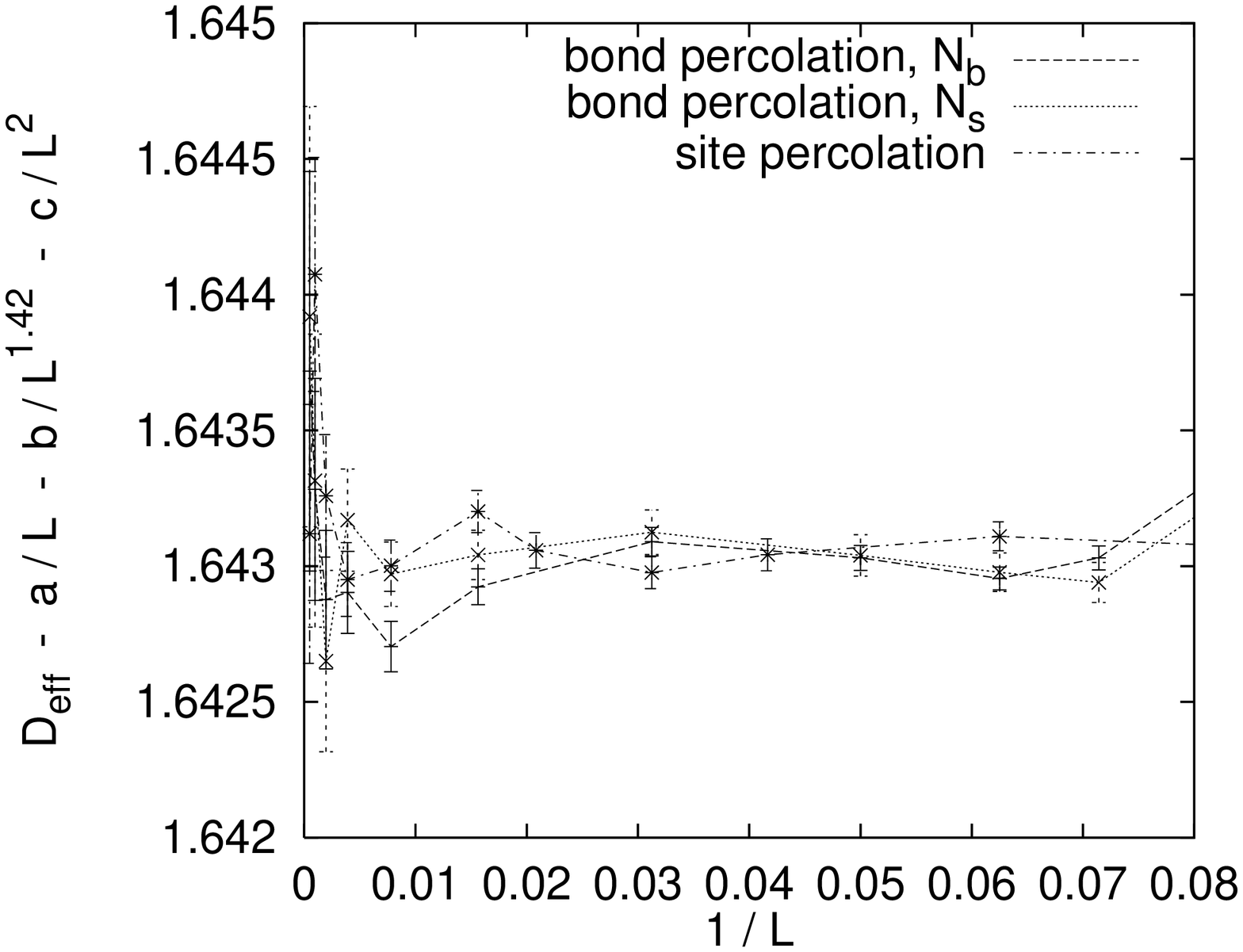,width=11cm,angle=270}}
   \caption{ \small 
        Similar to fig.2, but with three correction terms subtracted from each 
        curve. Two of these terms are analytic (with exponents $-1$ and $-2$), the 
        third has a common exponent $\Delta = -1.42$.
   }
   \label{fig3}
\end{figure}

If the above procedure were physically meaningful, the three curves in fig.2 
should be horizontal and should collapse. While the former is at least true for 
two of them (the curve for backbones defined via bonds in bond percolation, which 
should show cleanest behavior, is not flat), the latter is definitely not true. 
The subtraction reduced considerably the differences between the curves. 
But the error bars are so small that the remaining differences still
are highly significant. 

Indeed it is not surprising that the above procedure does not give satisfactory 
results. From universality one should have expected that $\Delta$ is the same 
for all three observables. Our findings that $\Delta$ varies between 
$0.67$ and $1.20$ shows that these can only be effective exponents, obtained by 
fitting by a single power a function which actually is a superposition of 
several powers. Unfortunately, trying to fit the data with two terms, 
$D_{b,{\rm eff}}(L) = D_b + a/L^{\Delta_1}+b/L^{\Delta_2}$ gives either no 
satisfactory fits (if too many values of $L$ are used in the fit) or ambiguous 
fits (if only very large values of $L$ are used). 

Our final procedure was the following. First of all, the bond percolation data 
strongly suggest that the leading non-analytic corrections have exponent $>1$. 
The leading term is then analytic, $\sim 1/L$, and the third important term is 
again analytic, $\sim 1/L^2$. We fitted the data therefore with ansatzes 
\be
   D_{b,{\rm eff}}(L) = D_b + a/L + b/L^\Delta + c/L^2, 
\ee
with {\it the same} exponent $\Delta$ for all three data sets. The fit (using 
all $L>10$) gave $\Delta = 1.42$. We do not quote any error bars since this 
value might change considerably if we include even more terms, or if we 
change the range of $L$ used for the fit. Also the constants $a, b,$ and $c$ 
should not be meaningful. But we checked that the value of $D_b$ obtained in 
this way is very robust and agrees for all three data sets, see fig.3.
Figure 3 shows some deviations from horizontal lines for very large $L$, 
but they seem to be statistical fluctuations. Our final estimate is 
\be
   D_b = 1.6432\pm 0.0008 \;.
\ee
The error bars are subjective. They were estimated by comparing with similar 
figures obtained by changing the range of $L$ and including a fourth term with 
$\Delta = 2.5$, but restricting its amplitude to a value $\leq |c|$.

Our estimate is five times more precise than that of 
\cite{grass}, but agrees with it within the error bars. 
It is of course completely incompatible 
with the prediction of \cite{larsson,saleur}, but it disagrees also with 
Huber's prediction by about 3 standard deviations. It is just outside the 
error bars of the estimate $1.650\pm 0.005$ of \cite{moukarzel}.

\section{Conductivity}

For conductivity measurements we used the Lobb-Frank algorithm, after removing all 
sites resp. bonds not belonging to the backbone. Since implementation of this algorithm 
is not quite trivial, and since our results were rather surprising, we performed 
very extensive tests. We compared the results of the Lobb-Frank algorithm with 
several other algorithms. In particular we used several variants of 
Fogelholm's method \cite{fogelholm} and the direct solution of the Kirchhoff 
equations. In all cases we compared also the conductivity of the backbone with 
that of the entire lattice, finding perfect agreement within the numerical 
accuracy. During these checks we also verified that the Lobb-Frank algorithm 
is by far the fastest.

In the Lobb-Frank algorithm one imagines wires to be attached to the lower left 
and upper right corners of the lattice. One then uses alternatingly the star-triangle and the 
triangle-star relations to ``push" the upper and left boundary of the lattice 
downward and to the right in such a way that the total conductivity is unchanged. 
When the calculation is finished, the whole lattice consists of only 
the lower and right boundaries. The lower boundary will be superconducting, 
and the entire resistance comes from the right boundary which consists of a string 
of resistors connected in series.

When removing parts that do not belong to the backbone, one still has to push the 
boundary through the entire lattice, whence the complexity cannot be smaller than 
$O(L^2)$. But most of the star-triangle and triangle-star relations will be trivial, 
leading to the substantial numerical improvement mentioned above. For bond percolation 
on a lattice with $L=4096$, the average CPU time on an DEC Alpha with 433 MHz was 
ca 16 sec per lattice. This included the determination of the backbone and, if this 
was non-zero, the computation of the conductivity.

\begin{figure}[ht]
   \centerline{\psfig{file=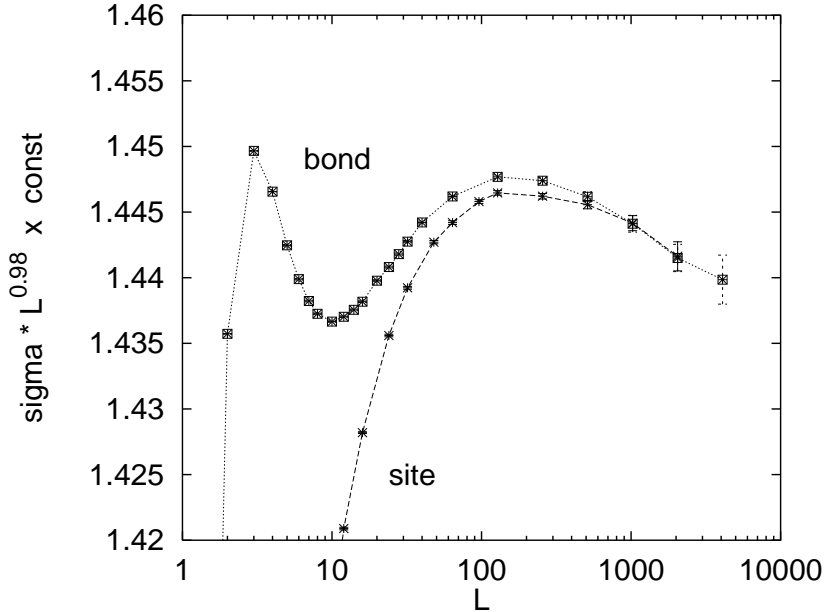,width=11cm,angle=270}}
   \caption{ \small 
        Conductivities on lattices of size $L\times L$, multiplied by $L^0.98$ and 
        by an arbitrary constant. The upper curve is for bond, the lower for site 
        percolation. The bond percolation data have been multiplied by a factor 
        1.2845 in order to coincide for large $L$ with the site percolation data.
   }
   \label{fig4}
\end{figure}

For bond percolation we computed the conductivity for all configurations. For 
site percolation we did it only for about half of the configurations, except that 
we did not compute conductivities for site percolation lattices with $L=4096$. In all
cases we produced data only at the exact (resp. numerically precise) percolation 
threshold. 

Altogether our sample involved more than $10^{14}$ wetted sites. This 
should be compared to ca. $10^{11}$ wetted sites in \cite{normand}. In 
addition, most of our data were obtained for lattices with much larger (transverse) 
size than those of \cite{normand}, and the algorithm used in \cite{normand} 
scales very unfavorably with lattice size. It should finally be pointed out that the 
Lobb-Frank algorithm can also be used (with slight modifications) for the strip 
geometry used in \cite{normand}, provided the periodic lateral boundary conditions are replaced 
by helical ones.
 
Results are shown in fig.4. In this figure we present both site and bond percolation 
conductivities, both multiplied by $L^{0.98}$. On the $x$-axis is plotted $\log L$. While 
we see a single-humped curve for site percolation, the data for bond percolation (which 
should be more clean according to the theoretical arguments given in sec.1) show an 
additional bump with maximum at $L=3$. For large $L$ both curves seem to converge, suggesting that 
the leading correction to scaling term is the same for bond and site percolation. 

\begin{figure}[ht]
   \centerline{\psfig{file=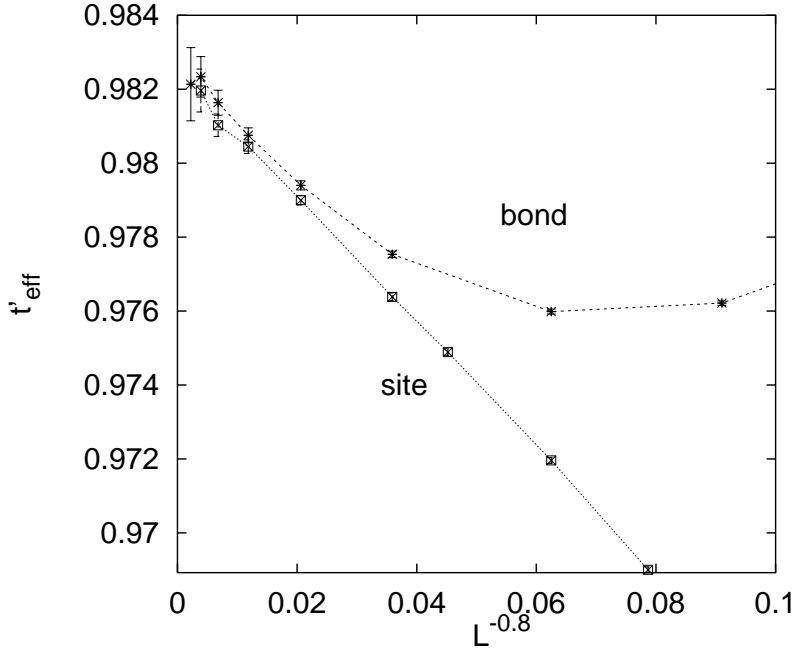,width=11cm,angle=270}}
   \caption{ \small
        Effective exponents computed from the data shown in fig.4, plotted against $1/L^{0.8}$.
   }
   \label{fig5}
\end{figure}

The bump seen for bond percolation at small $L$ is definitely not a statistical fluctuation 
(the error bars for $L<50$ are much smaller that the sizes of the points). 
It is not due to a programming error either, as seen by comparing with the data of 
\cite{lobb,frank}, and with unpublished data by D.J. Frank \cite{frank-unpub}
(for $L\leq 4$ we also computed $\sigma$ by exact enumerations, and obtained 
results in perfect agreement with the simulations). 
It is completely unexpected, and its origin is theoretically not understood. 

The bond percolation 
data shown in fig.4 could equally well be fitted by a log-periodic oscillation, 
\be
   \sigma = L^{0.98} (a + b \sin (\alpha \log (L/L_0))) \;.
\ee
It is mainly the good agreement with site percolation at large $L$ which makes
such an extrapolation very unlikely.

The best estimate for the exponent $t'$ is obtained by again defining an effective 
exponent $t'_{\rm eff}(L)$ by means of eq.(\ref{deff}), and plotting it against a
suitable power of $1/L$. If there is a single dominant correction to scaling term, 
we obtain a straight line extrapolating to $t'$. Such a straight line cannot 
of course be expected for the bond percolation data, in view of the oscillation seen 
in fig.4. But the site percolation data do show a beautifully straight line when plotted 
against $1/L^{0.8}$, see fig.5, suggesting $\Delta = 0.8$. 
This should of course not be taken too serious in view of our experience with the 
backbone dimension. Therefore we base our final estimate of $t'$ not only on the 
straight line extrapolations in fig.5, but also on similar extrapolations for plots 
with $\Delta$ ranging from $0.7$ to $1.3$, and on fits with up to three correction 
terms with the same exponents as in the last section. The error bars in our result 
\be
   t' = 0.9825 \pm 0.0008 \;,  \qquad t = t'\nu = 1.3100\pm 0.0011\;.
\ee
include the associated uncertainty.

This estimate is nearly by a factor 2 more precise than the estimate of \cite{normand}, 
which itself was by far the most precise previous estimate. It is barely compatible 
with it. It is compatible with the estimate of \cite{frank}. It excludes the 
Alexander-Orbach conjecture, the conjecture $t'=1$ of \cite{levinstein}, and the 
recent prediction $t'=0.995\pm 0.001$ of \cite{milovanov}. When translated into 
exponents for diffusion in disordered lattices \cite{havlin}, it gives exponents which 
are more precise than all previous estimates. For random walks on the infinite incipient 
cluster it gives e.g. the fractal dimension \cite{havlin}
\be
   d_w = 2 + {t-\beta \over \nu} = 2.8784\pm 0.0008 \;.
\ee
Direct estimates of $d_w$ (not using the Einstein relation) have errors which are larger 
by at least one order of magnitude \cite{havlin}.

\section{Conclusion}

We have presented high statistics simulations of 2-d percolation, using what we believe 
to be the fastest algorithms known at present. Our efforts were concentrated at 
``a-thermal" exponents which cannot be computed easily by mapping percolation onto 
the 1-component Potts model and using the standard machinery for spin models. 
We obtained what we believe to be the most precise estimates for the backbone dimension 
and for the conductivity exponent $t$. Indeed, our data sets are much larger than 
previous ones, but our estimates for critical exponents are not so much more precise 
because of very substantial corrections to scaling. 

For most observables we found that these corrections to scaling could not 
be described by a single term. In some cases (backbone dimension) a single term 
would superficially seem to give a decent fit, but such fits would be rather 
misleading. In that case it was the comparison between bond and site percolation 
which allowed us to reach a more definite and satisfactory conclusion. 

For conductivities, the main phenomenon was an unexpected anomaly for 
bond percolation which had a priori been expected to yield the cleanest signal. By itself, 
this anomaly could have been interpreted in various ways, including logperiodic 
oscillations. Again it was only the comparison with site percolation which ruled 
out such exotic possibilities, and which allowed us to obtain a precise estimate
of the critical exponent. 

Although the precise values of the critical exponents should be interesting by 
themselves, we believe that the more important message of the present paper is that 
one should be extremely careful with corrections to scaling. This is particularly true 
as data get more and 
more precise. A simple least square fit on a log-log plot might be appropriate for 
extremely crude data, although it is dangerous even then. Much better is of course the widely 
used strategy of making several least square fits, excluding more and more data points 
which might be outside the proper scaling region. An alternative is to fit ansatzes 
which contain suspected correction terms. This is certainly strongly advised, but 
as seen from the above examples it can also be very misleading, if the corrections 
to scaling have unexpected structures. The main point we want to stress is that 
the latter might be more common than is often appreciated.

\section*{Acknowledgements}

I am indebted to C. Moukarzel and D. Stauffer for very interesting discussions, to 
D. Frank and C. Lobb for correspondence and for showing me their unpublished data, 
to G. Huber for correspondence, discussions, and for allowing me to quote his prediction, 
and to R. Ziff for pointing out Ref. \cite{aharoni}.

\newpage
\eject

\newpage

\begin{table}
\footnotesize
\begin{center}
\begin{tabular}{|c|c|c|c|c|c|c|c|}
\hline
  $L$ & nr. of conf. & $P_{\rm span}$ & $N$ & $\Delta N/N$ & nr. of conf. & CPU & $\sigma $  \\ \hline
    2 & 143000000  &  0.579216 &      3.0119 & 0.21368 &   93000000  &      0.32  &  0.62607591(2657)  \\
    3 & 130000000  &  0.566722 &      5.7037 & 0.25652 &   50000000  &      0.30  &  0.44736350(3025)  \\
    4 & 101400000  &  0.555578 &      9.0238 & 0.26697 &   77000000  &      0.65  &  0.34771594(2006)  \\
    6 & 146000000  &  0.541412 &     17.2180 & 0.27888 &  116000000  &      2.35  &  0.23998530(1185)  \\
    8 & 108600000  &  0.532984 &     27.2498 & 0.28513 &   77100000  &      3.42  &  0.18313588(1134)  \\
   12 & 103400000  &  0.523401 &     52.1417 & 0.29161 &   63000000  &      6.99  &  0.12444040(0869)  \\
   16 & 152900000  &  0.518070 &     82.7607 & 0.29484 &  117900000  &     18.29  &  0.09435188(0486)  \\
   24 & 100000000  &  0.512537 &    159.1031 & 0.29815 &   65700000  &     20.75  &  0.06374185(0444)  \\
   32 & 142100000  &  0.509451 &    253.4148 & 0.29965 &   53800000  &     44.47  &  0.04820400(0371)  \\
   48 & 105150000  &  0.506488 &    489.3780 & 0.30101 &   44950000  &     93.54  &  0.03247599(0276)  \\
   64 &  94028000  &  0.504897 &    781.6404 & 0.30181 &   34493000  &    126.20  &  0.02452290(0238)  \\
   96 &  70956000  &  0.503185 &   1514.462  & 0.30239 &   34256000  &    260.82  &  0.01649989(0162)  \\
  128 &  65796500  &  0.502520 &   2423.569  & 0.30255 &   22015500  &    262.38  &  0.01245207(0152)  \\
  256 &  31548400  &  0.501203 &   7537.881  & 0.30289 &   15485600  &    374.04  &  0.00631185(0092)  \\
  512 &  12368400  &  0.500702 &  23493.92   & 0.30305 &    6688300  &    455.17  &  0.00319855(0071)  \\
 1024 &   5616010  &  0.500022 &  73290.45   & 0.30291 &    2136010  &    751.54  &  0.00162002(0064)  \\
 2048 &   1497050  &  0.500224 & 228958.62   & 0.30236 &     555400  &    729.32  &  0.00081988(0063)  \\
 4096 &    522500  &  0.500553 & 714407.33   & 0.30315 &          0  &    372.47  &      --- \\    
\hline
\end{tabular}
\end{center}
\caption{Raw data for site percolation. The second column contains the total number of configurations, while column 
nr. 6 gives the number of configurations for which also $\sigma$ was computed. Column nr. 5 gives the relative 
rms. width of the distribution of $N$, the error on $N$ is given by this number divided by the square root of 
the number of configurations and multiplied by $N$. CPU times are measured in hours, and are quoted for a 
433 MHz DEC Alpha workstation. The error on the spanning probability $P_{\rm span}$ is given by 
$\sqrt{N P_{\rm span} (1-P_{\rm span})}$.} 
\label{table1}
\bigskip
\bigskip
\bigskip
\bigskip
\vspace{5.cm}
\end{table}

\newpage

\begin{table}
\footnotesize
\begin{center}
\begin{tabular}{|c|c|c|c|c|c|c|c|c|}
\hline
  $L$ & nr. of conf. & $P_{\rm span}$ & $N_s$ & $\Delta N_s/N_s$ & $N_b$ & $\Delta N_b/N_b$ & CPU & $\sigma $  \\ \hline
    2 &         32 & 0.500000 &    1.5000 &  0.33333 &    2.9375 &  0.00000 &      0.00 &   0.56666666(0000) \\
    3 &       8192 & 0.500000 &    3.7708 &  0.33160 &    5.8502 &  0.37107 &      0.00 &   0.38455100(0000) \\
    4 &   33554432 & 0.500000 &    6.6766 &  0.33030 &    9.5497 &  0.36797 &      0.22 &   0.28945570(0000) \\
    5 &  130500000 & 0.499929 &   10.1512 &  0.32876 &   13.9466 &  0.36176 &      2.99 &   0.23194312(1112) \\
    6 &  169000000 & 0.500003 &   14.1503 &  0.32732 &   18.9736 &  0.35585 &      4.24 &   0.19364485(0824) \\
    7 &  142300000 & 0.499958 &   18.6394 &  0.32593 &   24.5949 &  0.35075 &      3.90 &   0.16630186(0775) \\
    8 &  163300000 & 0.500047 &   23.5926 &  0.32471 &   30.7701 &  0.34639 &      5.77 &   0.14580423(0637) \\
   10 &  211000000 & 0.500086 &   34.7923 &  0.32259 &   44.6763 &  0.33970 &     21.79 &   0.11711663(0452) \\
   12 &  155850000 & 0.499989 &   47.6161 &  0.32084 &   60.5351 &  0.33466 &     20.17 &   0.09797904(0442) \\
   14 &  162450000 & 0.499982 &   61.9535 &  0.31939 &   78.1942 &  0.33077 &     29.32 &   0.08427230(0373) \\
   16 &  155625000 & 0.499968 &   77.7141 &  0.31823 &   97.5718 &  0.32776 &     16.74 &   0.07396709(0335) \\
   20 &  111260000 & 0.500019 &  113.2612 &  0.31638 &  141.1415 &  0.32349 &     39.88 &   0.05950465(0320) \\
   24 &  126480000 & 0.499998 &  153.8188 &  0.31484 &  190.7419 &  0.32035 &     55.29 &   0.04980422(0252) \\
   28 &   86150000 & 0.500022 &  199.0584 &  0.31376 &  245.9541 &  0.31817 &     38.56 &   0.04285037(0263) \\
   32 &   90877000 & 0.500048 &  248.7624 &  0.31288 &  306.5485 &  0.31640 &     45.61 &   0.03761911(0225) \\
   40 &   62065000 & 0.499927 &  360.6193 &  0.31137 &  442.6715 &  0.31391 &     76.40 &   0.03026061(0219) \\
   64 &   46290000 & 0.499963 &  785.9377 &  0.30896 &  959.0467 &  0.31020 &    116.04 &   0.01911771(0161) \\
  128 &   30549000 & 0.499966 &  2467.540 &  0.30637 &  2995.707 &  0.30663 &    203.22 &   0.00970226(0101) \\
  256 &   14827100 & 0.500117 &  7724.715 &  0.30496 &  9354.184 &  0.30491 &    481.36 &   0.00491787(0074) \\
  512 &    4944350 & 0.500113 &  24158.26 &  0.30403 &  29215.70 &  0.30416 &    701.85 &   0.00249115(0065) \\
 1024 &    1654800 & 0.499833 &  75475.26 &  0.30433 &  91164.71 &  0.30394 &    768.83 &   0.00126117(0056) \\
 2048 &     647800 & 0.499102 &  235948.2 &  0.30314 &  285027.7 &  0.30349 &    711.75 &   0.00063823(0046) \\
 4096 &     198470 & 0.498635 &  737251.7 &  0.30367 &  890231.0 &  0.30373 &    852.14 &   0.00032320(0042) \\
\hline
\end{tabular}
\end{center}
\caption{Raw data for bond percolation. 
Column 8 gives total CPU times from runs on various Sun Ultra and DEC Alpha machines 
with different speeds. 
For $L\leq 4$, the second column shows the total 
number of distinct configurations, and all results were obtained by exact enumerations. 
For further explanations see table 1.}
\label{table2}
\end{table}

\end{document}